# Pre-Patterned Superconducting Contacts for Clean Superconductor–Topological Material Interfaces Enabling Long-Range Josephson Coupling

Yong-Bin Choi[1,2], Chang-Won Choi[3], Luke Holtzman[4], Hoil Kim[1], Seongwoo Kang[1], Kenji Watanabe[5], Takashi Taniguchi[5], James Hone[4], Jun Sung Kim[1,6], Si-Young Choi[3] and Gil-Ho Lee[1,†]

[1]Department of Physics, Pohang University of Science and Technology, Republic of Korea

[2]Max Planck POSTECH Center for Complex Phase of Materials, Pohang, Republic of Korea

[3]Department of Materials Science and Engineering, Pohang University of Science and Technology, Pohang, Republic of Korea

[4]Department of Applied Physics and Applied Mathematics, Columbia University, New York, NY, USA

[5]National Institute for Materials Science, Tsukuba, Japan

[6]Center for Artificial Low Dimensional Electronic Systems, Institute for Basic Science (IBS), Pohang, 37673 Republic of Korea

[†]Corresponding authors: lghman@postech.ac.kr (G.-H.L.)

**Phase-coherent superconducting proximity in topological materials requires clean superconductor–topological material (SC–TM) interfaces, yet conventional top-contact fabrication often degrades them through oxidation, polymer residue, and process-induced disorder. Here we introduce a pre-patterned superconducting bottom-contact architecture in which MoRe/Au electrodes are defined before van der Waals crystal transfer, thereby avoiding on-flake lithography after transfer. In $WTe_2$- and $Bi_{1.5}Sb_{0.5}Te_{1.7}Se_{1.3}$-based Josephson junctions, this architecture yields systematically larger $I_cR_N$ and longer-ranged coupling than conventional top contacts. Cross-sectional STEM/EDS reveals atomically abrupt, chemically well-separated interfaces. These results establish pre-patterned SC–TM contacts as a practical route to reproducible,**

**micrometer-scale Josephson platforms in van der Waals topological materials.**

Topological materials (TMs), particularly topological insulators (TIs), host gapless boundary states with linear dispersion. When proximitized by an *s*-wave superconductor, these systems have been predicted to provide a route toward topological superconductivity[1, 2] and associated Majorana-based device concepts. Proximity-induced superconducting transport through such boundary-localized channels is highly sensitive to contact quality and interfacial disorder. Developing a reproducible contact strategy that preserves a clean, well-defined SC–TM interface is therefore essential for reliable superconducting proximity devices in van der Waals (vdW) topological materials. Despite rapid progress in ultraclean vdW electronics, reproducibly achieving highly transparent superconducting contacts to TMs remains a major bottleneck for reliable transport with proximity-superconductivity. This limitation primarily arises from the difficulty of preserving a clean superconductor–topological material (SC–TM) interface in the presence of oxidation, polymer residues, and process-induced disorder, particularly in air-sensitive vdW materials.

For non-superconducting vdW devices, fabrication strategies that preserve the intrinsic electronic quality of the materials are well established. High mobility in vdW materials has been achieved using techniques such as hexagonal boron nitride (hBN) encapsulation[3, 4], residue-free transfer[5, 6], and pre-patterned electrodes[7-9]. In parallel, substantial effort has focused on improving electrical contacts between vdW materials and electrodes, through contact-transfer techniques[10-13] or the careful selection of adhesion layers[14-16]. In contrast, systematic and reproducible routes to atomically clean and highly transparent SC–TM interfaces remain limited, as many superconducting materials are prone to rapid oxidation upon exposure to air. Notable approaches include in-situ amorphous Te capping[17-19] and hBN encapsulation[20-22], which protect pristine surfaces; however, opening vias to define metal electrodes necessarily removes the capping layers locally, exposing the surface of TM with the risk of oxidation or contamination during etching of capping layers as well as before electrode deposition. Recent work has achieved transparent SC–TM interfaces with $XTe_2$ (X=W, Mo) by using Pd diffusion to create a superconducting $PdTe_x$ interlayer[23-28] as demonstrated by $WTe_2$–$PdTe_x$ Josephson junctions (JJs)[23-26]. While promising, Pd-diffusion-based contacts rely on forming a superconducting PdTex layer whose thickness and lateral extent are difficult to control.

Here, we introduce a pre-patterned superconducting bottom-contact architecture that preserves a clean SC–TM interface by avoiding resist-based lithography and wet processing on the material surface, while maintaining an atomically abrupt interface and a lithographically well-defined contact footprint. This approach also enables deterministic control of junction geometry for reproducible device fabrication. To demonstrate its effectiveness, we employ the geometry of a Josephson junction (JJ), in which two superconductors are coupled through a weak link, allowing us to probe proximity-induced superconductivity in TMs. Josephson coupling is identified by magnetic-field interference (Fraunhofer patterns). A key figure of merit is the characteristic voltage $I_c R_N$, where $I_c$ is the critical current and $R_N$ is the normal-state resistance of the junction. Using this metric, we benchmark proximity coupling across junction lengths and device architectures.

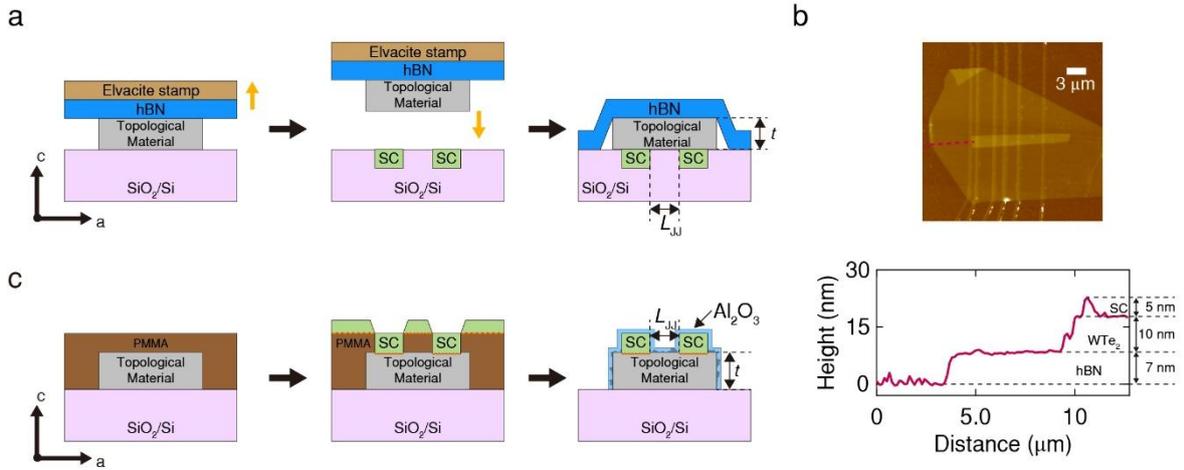

**Figure 1. Pre-patterned superconducting contacts and conventional top contacts. a,** Schematic illustration of the pre-patterned superconducting (SC) contact approach, where the TM flake is transferred onto pre-defined SC electrodes using an hBN/Elvacite stamp. $L_{JJ}$ and $t$ denote the junction length and flake thickness, respectively. **b,** Atomic force microscopy (AFM) image of an hBN/WTe$_2$ device and the corresponding height profile along the red dashed line, showing thicknesses of ~ 7 nm for hBN and ~10 nm for WTe$_2$, and an apparent contact step height of ~5 nm relative to the SiO$_2$ surface. **c,** Conventional top-contact fabrication, where SC electrodes are defined directly on the TM surface using a poly(methyl methacrylate) (PMMA) resist and lift-off lithography.

To address the challenge of forming clean superconducting contacts to vdW topological materials, we develop a pre-patterned superconducting bottom-contact architecture. In this approach, superconducting electrodes are first defined on the substrate prior to vdW material transfer, allowing the vdW materials to be placed directly onto pre-formed contacts while avoiding resist-based lithography and wet processing on the material surface. This geometry preserves a clean SC–TM interface while retaining lithographic control of the junction geometry (Fig. 1a).

Electrodes were recessed into the $SiO_2$/Si substrate to minimize the contact-edge step height (~1–5 nm), producing a nearly planar electrode surface. The electrodes consist of a MoRe superconducting layer (36 nm) capped with a thin Au layer (~5 nm), which protects MoRe from oxidation while remaining thin enough to be proximitized by the underlying superconducting layer. An hBN/TM stack was then dry-transferred onto the pre-patterned electrodes, ensuring that the junction region remains free from on-flake lithography. Using this approach, we fabricated devices based on multilayer $WTe_2$ encapsulated by hBN (Fig. 1c), as well as $Bi_{1.5}Sb_{0.5}Te_{1.7}Se_{1.3}$ (BSTS), a bulk-insulating three-dimensional topological insulator, passivated with the Elvacite polymer, to test the generality of the contact scheme. (see Methods for fabrication detail).

For comparison, conventional top-contact devices were fabricated following previous procedures[29, 30] (Fig. 1b), where superconducting electrodes were defined directly on the TM using electron-beam lithography. Immediately prior to metal deposition, in-situ Ar-ion etching was used to remove the oxidized TM surface. For a fair comparison with the pre-patterned bottom-contact devices, we employed an Au interfacial layer with the same thickness (~5 nm), deposited on the TM surface, followed by deposition of a 45-nm-thick MoRe superconducting layer. Because the conventional devices are not hBN-encapsulated, the exposed TM surface was passivated to mitigate ambient degradation during subsequent handling, either with a thin $Al_2O_3$ layer deposited by atomic layer deposition (ALD) after electrode fabrication for $WTe_2$ devices.

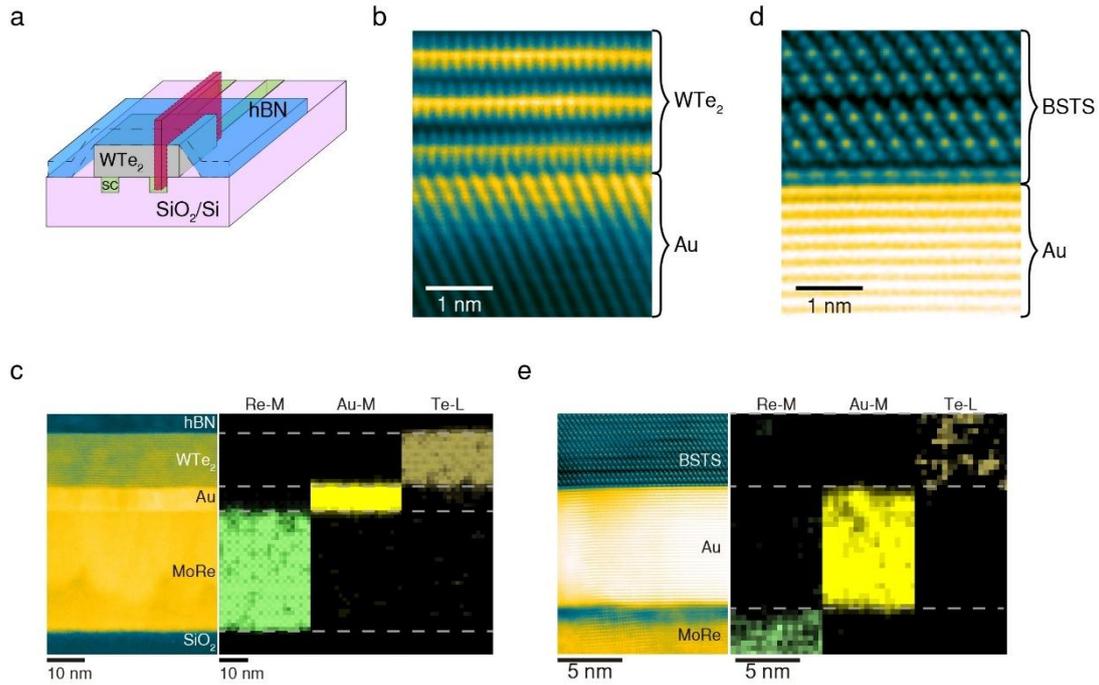

**Figure 2. Cross-sectional scanning transmission electron microscopy (STEM) and energy-dispersive X-ray spectroscopy (EDS) of the pre-patterned bottom-contacted (PB) devices. a,** Schematic of the PB-WTe$_2$ device. The red-shaded region indicates the lamella used for the cross-sectional STEM measurement. **b,** High-resolution STEM image of the interface between the WTe$_2$ crystal and the Au layer. **c,** False-colored STEM cross-section of the hBN/WTe$_2$/Au/MoRe stack and corresponding EDS elemental maps (Re-M, Au-M, and Te-L) acquired from the same area. Gray dashed lines indicate approximate layer boundaries. **d,** High-resolution STEM image of the interface between the BSTS crystal and the Au layer. **e,** False-colored STEM cross-section of the BSTS/Au/MoRe stack and corresponding EDS elemental maps acquired from the same area. Gray dashed lines indicate approximate layer boundaries.

Hereafter, we denote the pre-patterned superconducting bottom-contact architecture as PB (e.g., PB-WTe$_2$ and PB-BSTS). To directly assess the structural and chemical quality of the contact interfaces, we prepared cross-sectional lamellae by focused-ion-beam (FIB) lift-out across the junction region of a PB-WTe$_2$ device (Fig. 2a). Atomic-resolution high-angle annular dark-field scanning transmission electron microscopy (STEM) shows that the WTe$_2$ lattice remains ordered up to the last layer adjacent to Au, followed by an abrupt contrast change to the metal (Fig. 2b). The dispersive X-ray spectroscopy (EDS) maps shown in Fig. 2c resolve the MoRe layer (Re-M), Au layer (Au-M), and WTe$_2$ layer (Te-L). Within our experimental sensitivity, these data indicate chemically distinct layers with no extended interdiffusion region or obvious amorphous reaction layer at the WTe$_2$/Au interface. In contrast, conventional top-contact devices fabricated directly on WTe$_2$ exhibit a rough, structurally ill-defined interface with evidence of process-induced damage and intermixing (Supporting Fig. S1). PB-BSTS cross-sections (Figs. 2d and e) show a similarly well-defined interface. The STEM image (Fig. 2d) and corresponding EDS maps (Fig. 2e) reveal an abrupt contrast change at the BSTS/Au interface, with no extended amorphous reaction layer or intermixing observed within our experimental sensitivity. Together, these results indicate that the PB contact geometry yields structurally and chemically well-defined TM/Au interfaces.

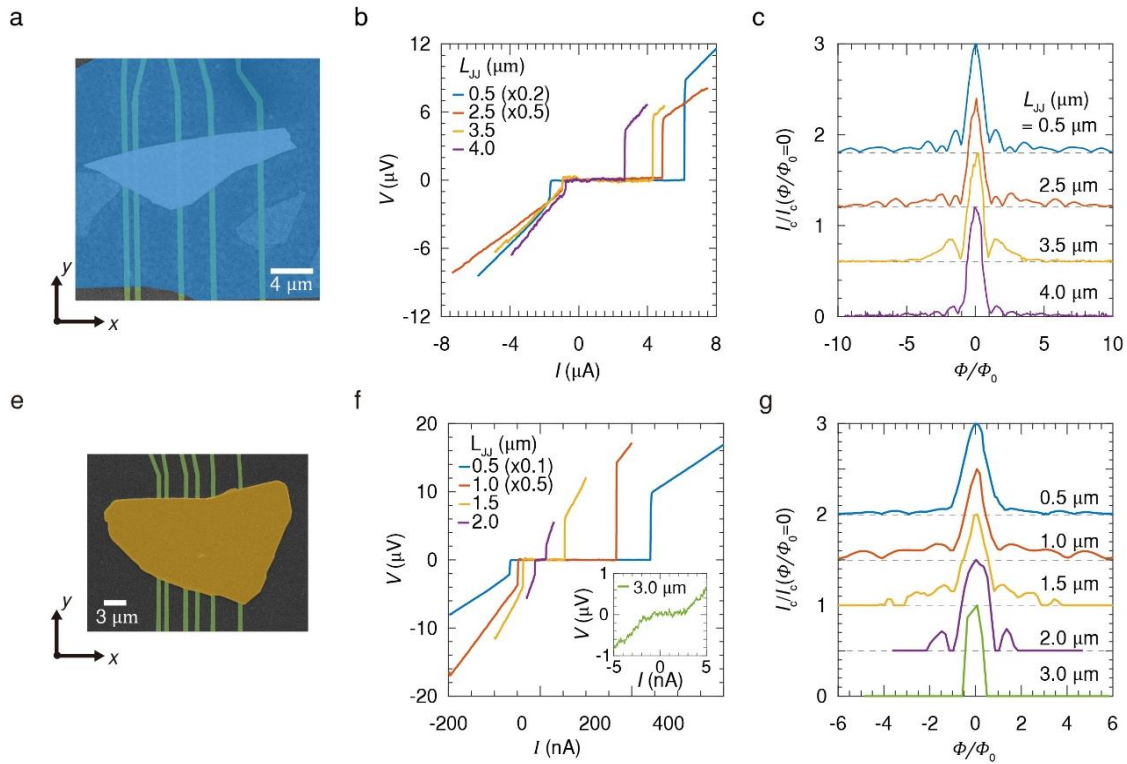

**Figure 3. Josephson junction measurements in pre-patterned bottom-contact (PB) devices. a, e,** False-colored scanning electron micrograph of the PB-WTe$_2$ (**a**) and PB-BSTS (**e**). Colors denote hBN (blue), WTe$_2$ (gray), BSTS (orange) and superconducting electrodes (green). **b, f,** Typical *I–V* characteristics measured by sweeping the bias current from negative to positive values at a base temperature of 20 mK for different junction length ($L_{JJ}$) in PB-WTe$_2$ (**b**) and PB-BSTS (**f**). For clarity, the *I–V* traces for $L_{JJ}$ = 0.5 and 2.5 μm in PB-WTe$_2$ and $L_{JJ}$ = 0.5

and 1.0 μm in PB-BSTS are rescaled on both axes. **c, g,** Magnetic-field interference patterns of the normalized critical current $I_c/I_c(\Phi/\Phi_0 = 0)$ as a function of magnetic flux $\Phi$ normalized by the superconducting flux quantum $\Phi_0 = h/2e$ for PB-WTe$_2$ (**c**) and PB-BSTS (**g**). Traces are vertically shifted for clarity.

We next examine Josephson transport in the PB devices through magnetic-field interference. PB-WTe$_2$ with junction lengths $L_{JJ}$ ranging from 0.5 μm to 4.0 μm shown in Fig. 3a were investigated (see Supporting Information Tables S1-S4 for full device information). Representative I–V characteristics measured at a base temperature of 20 mK are shown in Fig. 3b. For the shortest junction ($L_{JJ} = 0.5$ μm), we obtain $I_c = 30.7$ μA and $R_N = 1.54$ Ω. As $L_{JJ}$ increases, $I_c$ decreases while $R_N$ slightly increases. Notably, even the longest junction ($L_{JJ} = 4.0$ μm) exhibits a clear supercurrent with $I_c = 2.65$ μA and $R_N = 1.92$ Ω, corresponding to $I_c R_N = 5.09$ μV. Magnetic-field ($B$) dependence of $I_c$ exhibits a Fraunhofer interference pattern for all junctions (Fig. 3c and Supporting Information Figure S2). The oscillation period evolves from $\Delta B = 6.5$ G for the shortest junction to $\Delta B = 1.3$ G for the longest junction. These values are consistent with one flux quantum $\Phi_0$ threading an effective junction area $WL_{eff}$, where $W$ is the junction width, $L_{eff} = L_{JJ} + 2L'$ and $L' \approx$ 250nm, comparable to half the width of the superconducting electrode [29, 31, 32].

We now present similar measurements for PB-BSTS devices (Fig. 3e). Junctions with lengths ranging from $L_{JJ} = 0.5$ μm to 3.0 μm were fabricated. For the shortest junction ($L_{JJ} = 0.5$ μm), we obtain $I_c = 2.34$ μA and $R_N = 46.1$ Ω, while the longest junction ($L_{JJ} = 3.0$ μm) shown in the inset of Fig. 3f still shows a finite supercurrent with $I_c = 2.1$ nA and $R_N = 344$ Ω, corresponding to $I_c R_N = 0.72$ μV (Fig. 3f). The Fraunhofer interference patterns were also observed (Fig. 3g and Supporting Information Figure S3).

Together, these measurements demonstrate Josephson coupling over junction lengths up to 4 μm in WTe$_2$ and 3 μm in BSTS. Combined with the structural characterization in Fig. 2, these results show that the pre-patterned TM/Au/MoRe contacts support robust proximity Josephson coupling across topological materials.

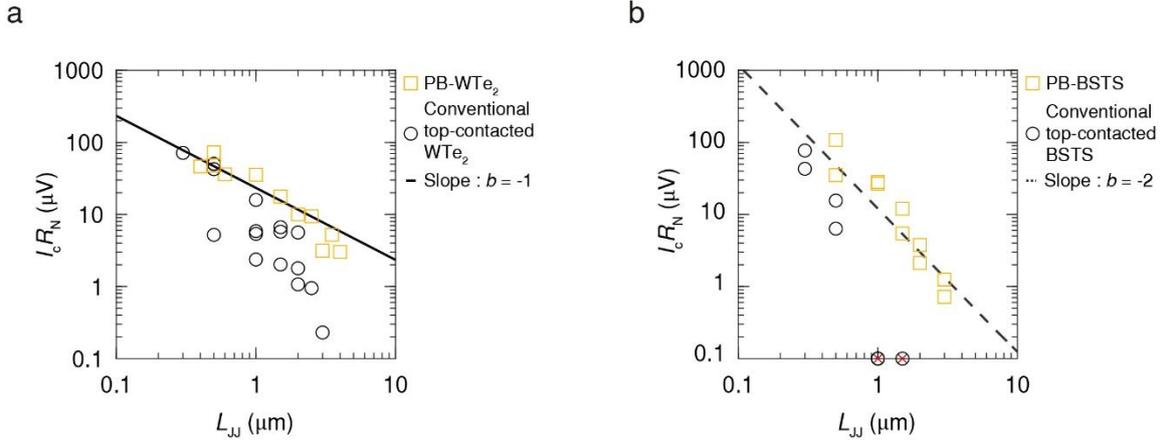

**Figure 4. Length dependence of Josephson coupling in pre-patterned and conventional devices. a**, **b**, The measured $I_cR_N$ is plotted versus $L_{JJ}$ for pre-patterned bottom-contact (PB) devices (yellow squares) and conventional top-contacted devices (black circles) based on WTe$_2$ (**a**) and BSTS (**b**). The lines correspond to a power-law scaling $I_cR_N \propto L_{JJ}^b$ with $b = -1$ for solid line and $b = -2$ for dashed line. Symbols marked with red crosses denote junctions in which no Josephson coupling was detected within our resolution in **b**.

We next benchmark the length dependence of Josephson coupling using the characteristic voltage $I_cR_N$, comparing PB devices based on WTe$_2$ (Fig. 4a) and BSTS (Fig. 4b) with conventional top-contact devices (Supporting Information Figures S4 and S5). Plotting $I_cR_N$ versus $L_{JJ}$ on log-log axes reveals a clear difference between the two contact geometries. In WTe$_2$, PB devices exhibit systematically larger $I_cR_N$ at comparable $L_{JJ}$, while conventional devices show a much stronger suppression with increasing length. The PB-WTe$_2$ data approximately follow an inverse-length scaling ($I_cR_N \propto L_{JJ}^{-1}$) up to $L_{JJ} \approx 2$ μm.

The superconducting coherence length is estimated as $\xi = \hbar v_F/2\Delta_{\text{MoRe}} \approx 58$ nm along the $a$-axis, with Fermi velocity in WTe$_2$ $v_F = 2.0 \times 10^5$ m/s (Ref. [33]) and superconducting gap of MoRe superconducting electrode $\Delta_{\text{MoRe}} = 1.14$ meV derived from its critical temperature 7.5 K. Since $L_{JJ} \gg \xi$ for all junctions studied here, the devices operate in the long-junction limit. The enhanced $I_cR_N$ in PB devices compared to conventional top-contact devices indicates more efficient proximity coupling. Consistently, cross-sectional STEM of conventional top-contacted WTe$_2$ devices (Supporting Information Figure S1) reveals a damaged interfacial layer formed due to direct metal deposition, which likely reduces contact transparency and weakens proximity coupling. A similar trend is observed in BSTS devices (Fig. 4b). PB-BSTS junctions maintain larger $I_c R_N$ over the measured length range and retain finite supercurrent up to $L_{JJ} = 3\ \mu m$, whereas conventional controls lose detectable supercurrent above $L_{JJ} \approx 1\ \mu m$. The PB-BSTS data follows a steeper scaling of $I_c R_N \propto L_{JJ}^{-2}$.

These scaling exponents of -1 for PB-WTe₂ and -2 for PB-BSTS are consistent with long ballistic and long diffusive Josephson coupling, respectively[34]. To further assess this interpretation, we analyze the length dependence of the normal-state resistance (Supporting Information Figure S6). PB-BSTS devices show an approximately linear $R_N W$ on $L_{JJ}$, consistent with diffusive transport, whereas PB-WTe₂ devices exhibit only weak dependence, indicative of ballistic transport. This normal-state resistance analysis is consistent with the junction transport regime inferred from the scaling of $I_c R_N$. Together, these results show that preserving a clean SC–TM interface via pre-patterned bottom-contact enables robust Josephson coupling over micrometer-scale junction lengths for both WTe₂ and BSTS.

In this study, we demonstrate a pre-patterned superconducting bottom-contact architecture that preserves a clean SC–TM interface in van der Waals devices by eliminating resist-based lithography on the crystal surface after transfer. A key challenge in realizing pre-patterned superconducting contacts is that most superconductors readily oxidize in air. Here, a thin Au capping layer protects the MoRe superconducting layer during fabrication while remaining sufficiently thin to allow efficient proximity coupling to the TM. A direct comparison with conventional top-contact devices shows that the pre-patterned junctions maintain significantly larger characteristic voltages $I_c R_N$ and sustain Josephson coupling over substantially longer junction lengths—up to 4 μm in WTe₂ and 3 μm in BSTS. Cross-sectional STEM/EDS measurements reveal atomically abrupt and chemically well-defined SC–TM interfaces, while Fraunhofer interference patterns and Shapiro steps confirm phase-coherent proximity coupling. Together, these results show that preserving a clean SC–TM interface enables robust long-range Josephson coupling in air-sensitive van der Waals topological materials. An additional advantage of the pre-patterned approach is that air-sensitive TMs can be mechanically transferred onto fully fabricated superconducting contacts, allowing the device fabrication to be completed without exposing the TM surface to lithographic processing and thereby helping preserve the intrinsic material quality. Looking ahead, this platform enables next-generation experiments toward topological superconductivity, including multi-terminal junctions[1, 35-37], engineered topological heterostructures[38-41], and magnetic–superconducting hybrid platforms[42], that may previously have been obscured by imperfect contacts.

## Methods

**Device fabrication (pre-patterned bottom contacts)**. Pre-patterned superconducting electrodes were fabricated on a Si substrate with a 280-nm-thick $SiO_2$ layer using electron-beam lithography with a double-layer poly(methyl methacrylate) resist. To minimize the electrode-edge step height, shallow recessed trenches were defined by in situ Ar-ion milling ($SiO_2$ etch rate ≈ 10 nm/min). Without breaking vacuum, MoRe was sputtered in an Ar atmosphere at a deposition rate of 15 nm/min, followed by Au deposition at a base pressure below $2 \times 10^{-7}$ Torr. Lift-off was performed in acetone at 50 °C with sonication for 15 min. The electrode surface and edges were subsequently planarized by contact-mode AFM ironing (XE-7, Park Systems), which effectively removes sidewalls formed during metal lift-off, using a typical set-point force of 1000 nN.

**Dry transfer and encapsulation.** Exfoliation and dry transfer were carried out in an Ar-filled glovebox ($O_2$ and $H_2O$ < 0.1 ppm). hBN and $WTe_2$ flakes were sequentially picked up using an Elvacite stamp and transferred onto the pre-patterned electrodes to form hBN/$WTe_2$ stacks. The thickness of $WTe_2$ flakes was confirmed by AFM. For BSTS devices, hBN encapsulation was not used due to the larger crystal thickness (>80 nm); instead, the BSTS surface was protected by an Elvacite polymer layer.

**STEM/EDS characterization.** High-angle annular dark-field scanning transmission electron microscopy (HAADF-STEM) imaging and energy-dispersive X-ray spectroscopy (EDS) were performed using an aberration-corrected microscope (JEM-ARM200F, JEOL) equipped with a dual EDS system (JED-2300, JEOL; effective detection area 100 $mm^2$) at the Materials Imaging & Analysis Center, POSTECH. HAADF-STEM images were acquired at an accelerating voltage of 80 kV with a probe convergence semi-angle of 28 mrad and collection angles of 54–216 mrad. EDS maps were obtained in STEM mode with an energy dispersion of 0.01 keV per channel and a spatial resolution of 128 × 128 pixels.

**Acknowledgement**

Half of this research was supported by the Nano and Material Technology Development Program through the National Research Foundation of Korea (NRF), funded by the Ministry of Science and ICT (RS-2024-00444725). Additional support was provided by other NRF grants (RS-2022-NR068223, RS-2024-00393599, RS-2024-00442710, RS-2025-02317602) and the ITRC program (IITP-2025-RS-2022-00164799), also funded by the Ministry of Science and ICT. Further support was received from Agency for Defense Development funded by Defense Acquisition Program Administration (DAPA) (UI257011TE), and Samsung Science and Technology Foundation (SSTF-BA2101-06, SSTF-BA2401-03). J.S.K. acknowledges support from the NRF grants (RS-2024-00399173).

**Author Contributions**

G.-H.L. conceived and supervised the project. Y.-B.C. fabricated the devices and performed the transport measurements. S.K. contributed to device fabrication. C.-W.C. performed the STEM and EDX imaging under the supervision of S.-Y.C. L.H. grew $WTe_2$ under the supervision of J.H. H.K. grew BSTS crystals under the supervision of J.S.K. K.W. and T.T. provided the hBN crystals. Y.-B.C. and G.-H.L. analyzed the data and wrote the manuscript.